\documentclass[arXiv,onecolumn,
superscriptaddress]{revtex4}
\usepackage[intlimits]{amsmath}
\usepackage{amsfonts}
\usepackage{graphics}
\usepackage{bbm}
\usepackage{subfigure}
\usepackage[usenames]{color}
\usepackage{epstopdf,epsfig}
\usepackage{mathtools}
\usepackage{ulem}
\usepackage{soul}

\DeclarePairedDelimiter\ket{\lvert}{\rangle}
\DeclarePairedDelimiterX\braket[2]{\langle}{\rangle}{#1 \delimsize\vert #2}

\usepackage{indentfirst}

\usepackage{graphicx}
\usepackage{amsmath}
\usepackage{amsfonts}
\newcommand\be            {\begin{equation}}
\newcommand\bea           {\begin{equation}\begin{array}l\displaystyle}
\newcommand\ee            {\end{equation}}
\newcommand\bes           {\begin{subequations}}
\newcommand\esu           {\end{subequations}}

\newcommand\p            {\partial}

\def\3pt#1#2#3{{\langle{#1}\vert{#2}\vert{#3}\rangle}}
\def\barray{\begin{eqnarray}}
\def\earray{\end{eqnarray}}

\usepackage{tikz}
\usetikzlibrary{calc}
\newcommand*\circled[1]{\tikz[baseline=(char.base)]{
        \node[shape=circle,draw,minimum size=5mm, inner sep=0pt] (char)
        {\rule[-3pt]{0pt}{\dimexpr2.4ex+2pt}#1};}}

\begin{document}

\title{Integrable Floquet Hamiltonian for a Periodically Tilted $1D$ Gas 
}

\author{A. Colcelli}
\affiliation{SISSA and INFN, Sezione di Trieste, Via Bonomea 265, I-34136 
Trieste, Italy}

\author{G. Mussardo}
\affiliation{SISSA and INFN, Sezione di Trieste, Via Bonomea 265, I-34136 
Trieste, Italy}

\author{G. Sierra}
\affiliation{Instituto de F\'isica Te\'orica, UAM/CSIC, Universidad 
Aut\'onoma de Madrid, Madrid, Spain}

\author{A. Trombettoni}
\affiliation{CNR-IOM DEMOCRITOS Simulation Center, Via Bonomea 265, I-34136 
Trieste, Italy} 
\affiliation{SISSA and INFN, Sezione di Trieste, Via Bonomea 265, I-34136
Trieste, Italy}

\begin{abstract}
An integrable model subjected to a periodic driving gives rise generally to a non-integrable Floquet Hamiltonian. 
Here we show that the Floquet Hamiltonian of the integrable Lieb--Liniger model in presence of a linear potential 
with a periodic time--dependent strength is instead integrable and its quasi-energies can be determined using the Bethe ansatz approach. 
We discuss various aspects of the dynamics of the system at stroboscopic times and we also propose a possible experimental realisation of the periodically driven tilting in terms of  
a shaken rotated ring potential. 
\end{abstract}

\maketitle

\section{Introduction}

Floquet theory is a powerful tool to study differential equations whose coefficients are time--periodic functions \cite{Floquet883} and, for this reason, it is widely used in Quantum Mechanics in presence of a time--periodic Hamiltonian \cite{Shirley65,Grifoni98}. 
If the system is initially in state $\chi (t=0)$ and subjected to a periodic driving, then the Floquet Hamiltonian $H_F$ is the operator that formally 
gives the state of the system at multiples of the period $T$:
\be
\label{Floq_Ham_def}
\chi(t=nT)=e^{-i\frac{nT}{\hbar}H_F}\,\chi(t=0)\,.
\ee
$H_F$ depends on the parameters of the original driven Hamiltonian, $H$, 
and of course on the time--dependent perturbation. The eigenvalues of the Floquet Hamiltonian are the {\it quasi-energies} $\mathcal{E}_F$. 
Often referred as Floquet engineering \cite{Eckardt2017,Oka18}, time--periodic driving allows to construct interesting effective $H_F$ with novel physical properties as, for instance, dynamic localization effects 
\cite{Dunlap1986}, suppression of inter--well tunneling in 
a Bose condensate subjected to a strongly driven optical lattice 
\cite{Creffield2003,Eckardt2005,Lignier2007,Kierig2008,Eckardt2009,Sierra2015} 
(see \cite{Eckardt2017} for more references), 
topological Floquet phases \cite{Kitagawa2010,Lindner2011}, 
time crystals 
\cite{Wilczek2013,Watanabe2015,Choi2017,Zhang2017,Russomanno2017,Yao2017,SachaZak2018,Sacha2018}, 
dynamics in driven systems \cite{Russomanno2012,Goldman2014,Holthaus2016} and 
Floquet prethermalization \cite{Weidinger17,Herrmann18}. 

Within this framework, a natural question is whether one can have an integrable non-trivial Floquet Hamiltonian perturbing an interacting model with a time--periodic perturbation. Despite 
exactly solvable time--dependent Hamiltonians can be constructed \cite{Yuzbashyan18,Sinitsyn18}, the problem of finding an integrable Floquet Hamiltonian $H_F$ from undriven interacting (possibly integrable) Hamiltonian $H$ is in general a challenging one. Our scope is to present explicitly an interacting many-body system whose Floquet 
Hamiltonian is indeed integrable and therefore exactly solvable by means of integrability techniques \cite{Korepin1993,Mussardo10}. This implies, in particular, 
that we can have access to the exact spectrum of the quasi-energies 
and also to the wavefunctions of the system at stroboscopic times (multiples of the period of the driving). It is useful to stress that one of the difficulties in identifying integrable Floquet 
Hamiltonians is that the integrability of these Hamiltonians is not at all guaranteed by the integrability of the original time-independent undriven model
(see, for instance, \cite{Komnik16} where 
starting from the original BCS model the corresponding BCS gap equation in presence of a periodic driving 
is derived and solved numerically).  

The original time--independent system that we consider in the following is the Lieb--Liniger (LL) model which describes the $\delta$-contact repulsion between bosons in one dimension \cite{LiebLiniger1963}: this model is exactly solvable using Bethe ansatz techniques \cite{LiebLiniger1963,Yang1969}, and routinely used to describe (quasi) one--dimensional bosonic gases realized in ultracold atoms experiments (see the reviews \cite{Yurovsky2008,Bouchoule2009,Cazalilla2011}). Interestingly enough, this quantum model 
remains exactly solvable also when it is perturbed by an external time--periodic potential {\it linear}  
in the coordinates $x_i$ of the particles, as it happens for its classical counterpart given by the non--linear Schr\"odinger equation which remains solvable also in presence of an external linear time--dependent potential 
\cite{Chen,Ablowitz2004}. Our system is then the quantum analog of a classical mechanics 
problem consisting of a body (subjected to the gravity force) put on a slide which changes periodically its slope by rotating around a pin posed at the origin. 
Before addressing the many-body Floquet Hamiltonian of the LL model, it is interesting to discuss preliminarily two instructive cases of 1D systems subjected to time--periodic linear potential: (a) the case of one particle and (b) the case of two interacting particles.  

\section{One particle} 

The Schr\"odinger equation for a particle of mass $m$ in a linear time--periodic 
potential in $1D$ reads
\be
\label{onebody_schro}
i \hbar \frac{\p \chi}{\p t}=-\frac{\hbar^2}{2m} \frac{\p^2 \chi}{\p x^2} + x\,f(t)\,\chi(x,t)\,,
\ee
where $f(t)=f(t+T)$. Eq. (\ref{onebody_schro}) has been studied and solved in several different ways \cite{Berry1978,Rau1996,Guedes2001,Feng2001}. 
Here we derive and discuss the results coming from Floquet theory in order to study the behaviour of the system at stroboscopic times, i.e. at multiples of the 
period $T$. To this aim, we perform a gauge transformation on the wavefunction 
\be
\label{wavefunction_gaugetrasf}
\chi (x,t) = e^{i \theta(x,t)} \,\eta(y(t), t)\,,
\ee
where $y(t)=x-\xi(t)$, with $\xi(t)$ and $\theta(x,t)$ to be determined. Substituting (\ref{wavefunction_gaugetrasf}) 
into (\ref{onebody_schro}), we find that if $d\xi/dt= (\hbar/m) \p \theta/\p x$ and $-\hbar \p \theta/\p t = (\hbar^2/2m) 
(\p \theta/\p x)^2 + xf(t)$, then $\eta(y,t)$ 
satisfies the Schr\"odinger equation 
with no external potential in the new spatial variables,  
$i \hbar \frac{\p \eta}{\p t}= -\frac{\hbar^2}{2\,m} 
\frac{\p^2 \eta}{\p y^2}$. 
Once $\theta(x,t)$ has been fixed (see Appendix A), the solution of (\ref{onebody_schro}) comes from (\ref{wavefunction_gaugetrasf}) where $\eta(y,t)$ is readily determined from the free dynamics. 
To simplify the subsequent formulas let's consider here the condition
\be
\label{condition_f(t)}
\int_0^T f(\tau)\,d\tau=0\,,
\ee 
(referring to Appendix A when $\int_0^T f(\tau)\,d\tau \neq 0$). In this case the gauge phase $\theta$ at times multiple of $T$ does not depend any longer on the spatial variable, 
i.e. $\theta(x,nT) \equiv \theta(nT)$. 
We can now determine the Floquet Hamiltonian $H_F$ according to the definition (\ref{Floq_Ham_def}) and 
one gets 
\be
\label{Floq_HamG_onebody}
H_F=\frac{\hat{p}^2}{2m}+\hat{p} \frac{\xi(nT)}{nT}-\hbar \frac{\theta(nT)}{nT}\,,
\ee
where the term linear in the momentum operator 
determines the motion of the center of mass 
with constant velocity $\xi(nT)/nT$, as we're going to show. Notice that if the expectation value $\langle \hat{p}\rangle $ of the momentum vanishes at $t=0$, then at each stroboscopic time we have 
$\langle\hat{p}\rangle = 0$ as well, despite of the fact that the system moves between two successive stroboscopic times. 
This can be seen from Eq. (\ref{wavefunction_gaugetrasf}) since $\hat{p}$ commutes with the kinetic energy term, and given the fact that $\theta$ at stroboscopic times 
does not depend on $x$. More generally, the expectation value of $\hat{p}$ at stroboscopic times takes the initial $t=0$ value. 

Introducing now the periodic function $\mathcal{F}(t)$ such that $\frac{d\mathcal{F}}{dt}=f(t)$, with $\mathcal{F}(0)=0$, one finds
\be 
\label{theta_final_bis}
\theta(nT)=-\frac{I'}{2\,m\,\hbar}\,n\,, \,\,\,\,\,\,\,\,\,
\xi(nT)=-\frac{I}{m}\,n\,,
\ee
where $I \equiv \int_0^T \mathcal{F}(t) dt$ and $I'\equiv\int_0^T\mathcal{F}^2(t)dt$. 
Notice that the ratios $\frac{\xi(nT)}{nT}$ and $\frac{\theta(nT)}{nT}$ do  not depend on the time step $n$. 

One may also rewrite the Hamiltonian in (\ref{Floq_HamG_onebody}) as $H_F=[\hat{p}+m\xi(T)/T]^2/2m +C$, where $C$ is a constant, 
$C=-\hbar\theta(T)/T - (m/2) [\xi(T)/T]^2$, and applying the unitary transformation on the Hamiltonian 
$\tilde{H}_F \equiv e^{i a x/\hbar} H_F e^{-i a x/\hbar}$ with $a=m \xi(T)/T$, we finally get
$\tilde{H}_F=\frac{\hat{p}^2}{2m}+C$.

It is interesting to establish a connection between our system and the phenomena of suppression of tunneling for shaken periodic lattices \cite{Eckardt2017}, 
where a renormalization of the tunneling parameter [{\it i.e.} of the mass] is induced by the driving: to this aim, let's {\it formally} rewrite the 
Floquet Hamiltonian (\ref{Floq_HamG_onebody}) with a renormalized effective mass as  
\be
\label{Floq_HamA_onebody}
H_F\,=\frac{\hat{p}^2}{2\,m_{\rm eff}(\hat{p})}\,,
\ee
where the effective mass depends on the momentum operator as 
\be
\label{m_eff_1body}
\left[m_{\rm eff}(\hat{p})\right]^{-1}=\frac{1}{m}\mathbbm{\hat{1}}+\frac{2 \xi(T)}{T}\,\hat{p}^{-1}-2\hbar\frac{\theta(T)}{T}\hat{p}^{-2}\,,
\ee
$\mathbbm{\hat{1}}$ being the identity operator. 
Considering as wavefunction at the initial time a planewave 
$\chi(x,0) \propto e^{i\tilde{k}x}$, 
the effective mass depends on $\hbar\,\tilde{k}$, \textit{i.e.} 
the initial momentum of the particle and,  at stroboscopic times, the expectation value of the momentum operator is equal to 
the initial momentum $\left\langle\hat{p} \right\rangle(nT)=\hbar\,\tilde{k}$. As a particular example let's consider 
\be
\label{f(t)_sin}
f(t)=\ell\,\sin(\omega t)\,,
\ee
with $\omega=\frac{2\pi}{T}$ and $\ell$ is a parameter having dimension of energy divided by length. 
One gets
\be
\label{m_eff/m_onebody_f(t)}
\frac{m_{\rm eff}(\tilde{k})}{m} =\left[1-\frac{2\,\ell}{\hbar\,\omega\,\tilde{k}}+\frac{3}{2}\left(\frac{\ell}{\hbar\,\omega\,\tilde{k}}\right)^2 \right]^{-1}\,.
\ee
Notice that for $\ell/\hbar \omega \tilde{k} < 4/3$, then 
$m_{\rm eff}(\tilde{k})>m$. Therefore $m_{\rm eff}$ increases for large $\tilde{k}$: this behaviour has to be compared with the one of shaken lattices \cite{Eckardt2017}, 
where the effective mass always increases 
(and the momentum in the lattice is bounded to be in the Brillouin zone).

\section{Two particles}

Let us now consider a 1D system of two interacting particles in a linear time--periodic potential, described by the Schr\"odinger equation
\[
\label{Schro_twobodies_generic}
i\,\hbar\,\frac{\p \chi}{\p t}=\sum_{j=1}^2\Big(-\frac{\hbar^2}{2\,m} \frac{\p^2}{\p x_j^2}+V(x_j,t)\Big)\chi+V_{2b}(x_2-x_1) \chi\,,
\]
where $V_{2b}(x_2-x_1)$ is a generic interacting 
potential between the two particles, while 
$V(x_j,t)=x_jf(t)$
is the external oscillating potential (for the LL model one has $V_{2b}(x_2-x_1)=\lambda\,\delta(x_2-x_1)$, where $\lambda$ is the coupling strength \cite{LiebLiniger1963}).
To solve this  Schr\"odinger equation, we employ the same method discussed before: assuming once again 
the condition (\ref{condition_f(t)}), we perform the gauge transformation 
\be
\label{wavefunction_gaugetrasf_twobodies}
\chi (x_1,x_2,t) = \, e^{i \left[\theta(x_1,t)+\theta(x_2,t)\right]} 
\eta(y_1(t), y_2(t), t)\,,
\ee
where $y_j(t)=x_j-\xi(t)$. The wavefunction $\eta(y_1,y_2,t)$ satisfies the Schr\"odinger equation for two interacting particles but no longer 
with the external potential
\[
\label{twobodies_schro_eta}
i \hbar \frac{\p \eta}{\p t}= -\frac{\hbar^2}{2\,m} \left[\frac{\p^2 }{\p y_1^2}+\frac{\p^2 }{\p y_2^2}\right]\eta + V_{2b}(y_1 -y_2)\,\eta\,,
\]
while $\xi(t)$ and $\theta(x_j,t)$
obey Eqs. (\ref{conditions_integrab_app}) in Appendix A. The Floquet Hamiltonian, defined by the condition $\chi(x_1, x_2,nT)=e^{-i\frac{nT}{\hbar}H_F}\chi(x_1, x_2,t=0)$, is then 
\[
\label{Floq_HamG_twobodies}
H_F=\sum_{j=1}^2\left(\frac{\hat{p}_j^2}{2\,m}+\hat{p}_j\frac{\xi(T)}{T}-\hbar\frac{\theta(T) }{T}\right)+V_{2b}(x_2-x_1)\,.
\]

As an example, let's use this Floquet Hamiltonian to study the stroboscopic time evolution of a Gaussian wave-packet subject to a time periodic linear potential 
\be
\label{gauss_wp}
\chi(x_1,x_2,0)={\cal C} e^{-\left(x_1^2 +x_2^2\right)/2\sigma^2}\,,
\ee
where ${\cal C}$ is the normalization factor and the variance of the initial wavepacket is $\propto\sigma$. We choose $f(t)$ given by Eq. (\ref{f(t)_sin}), obeying the 
conditions (\ref{condition_f(t)}) 
and $f(n T) = 0$, for $n=0,1,\ldots$. While the details of the calculation are in Appendix B, here in Fig. \ref{fig1} we
plot  the diagonal density matrix $\rho(x,nT)=2\,\int_{-\infty}^\infty \left| \chi(x,x_2,nT)\right|^2\,dx_2$ at different stroboscopic times for several values of the coupling constant 
for $V(x)=\lambda \delta(x)$. We see that the center of mass moves with constant velocity $-\ell/m\,\omega$, in this case towards the left, 
 while the effect  of the interaction is to split  the wave-packet in two pieces. It is also interesting to study the time evolution of the variance of the coordinates $x_j$, given in  Fig. \ref{fig1}\circled{d}, where one can see that the higher the interaction strength, the more pronounced is the spreading of the wave-packet. In this case, such a spreading at stroboscopic times for finite coupling $\lambda$ can be well approximated as 
\[
\label{variance_check}
\Delta x_j (nT) \approx \frac{\sigma}{\sqrt{2}} \sqrt{1+\left(\frac{\hbar\,n T}{m\,\sigma^2}\right)^2\left(1+{\cal B}\, \frac{m 
\,\lambda \,\sigma}{2\,\hbar^2}\right)}\,,
\]
where ${\cal B} \approx 1.23$. Notice that for $\lambda=0$ one retrieves the expression of the spreading of the wavepacket in the free case, \textit{i.e.} $\Delta x_j^0 (nT) \approx \frac{\sigma}{\sqrt{2}} \sqrt{1+\left(\frac{\hbar\,n T}{m\,\sigma^2}\right)^2}$, while for $\lambda=\infty$, the Tonks-Girardeau gas, one has a divergent value for any $n$ since  the tail of the density matrix decay $\propto \/x^2$ even starting from a Gaussian.

\begin{figure}[t]
\hspace*{0.pt}\includegraphics[width=0.7\linewidth]{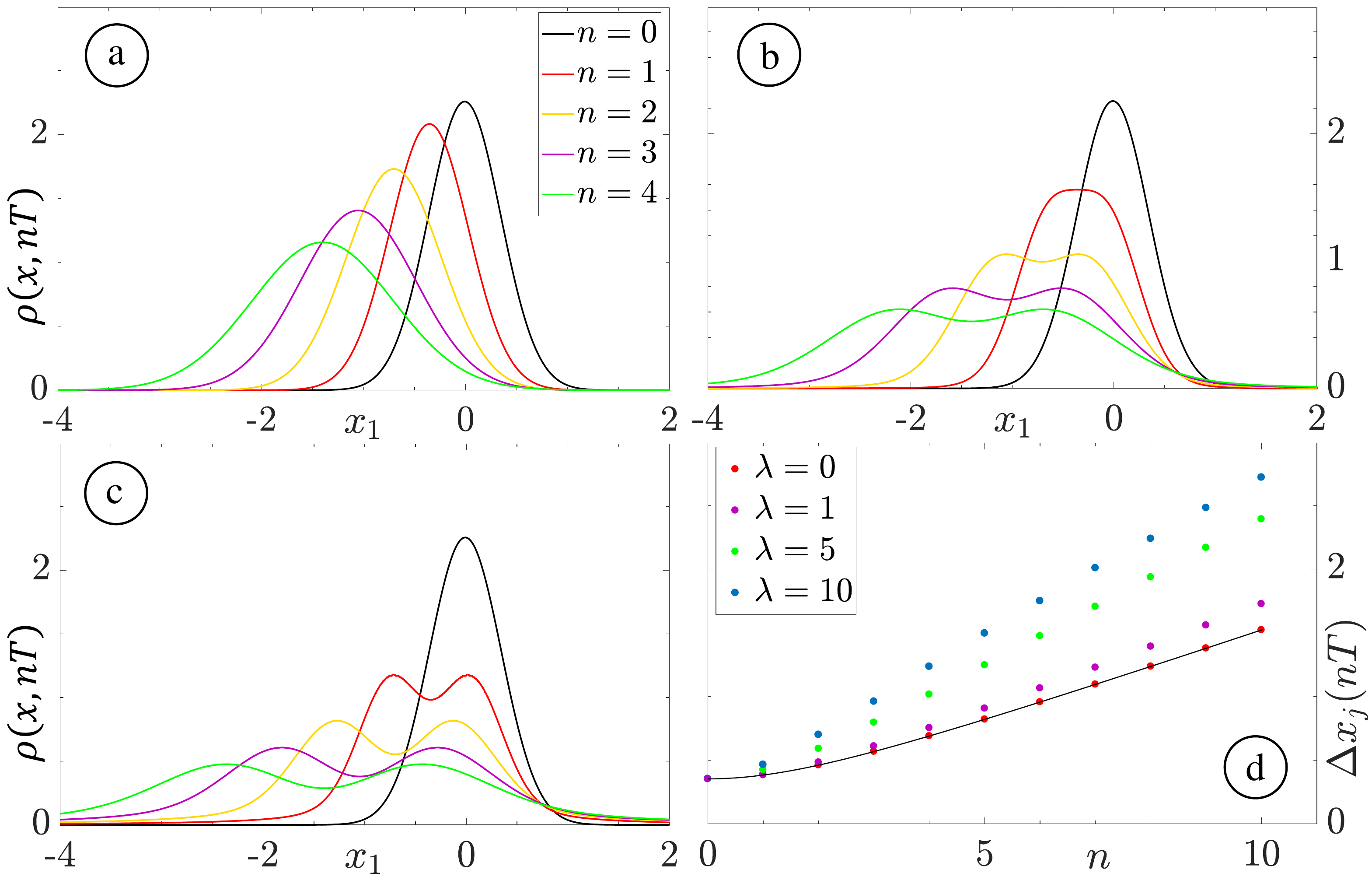}
\caption{\protect\circled{a}, \protect\circled{b}, \protect\circled{c}: Stroboscopic time evolution of the density matrix $\rho(x,nT)$ for a two--particles LL gas for different interacting strengths, starting from the Gaussian wavepacket (\ref{gauss_wp}) with dimensionless units chosen such that $\sigma=0.5$. \protect\circled{a}, \protect\circled{b}, \protect\circled{c} plots have $\lambda=0, 10, \infty$ respectively (moreover $T=\frac{2\pi}{60}$ and $\ell=200$). \protect\circled{d}: Time evolution of the variance at stroboscopic times, $\Delta x_j (nT)$, for different values of the interaction strength $\lambda$. The analytic behaviour for the free case, $\lambda=0$,  is reported as black solid line.}
\label{fig1}
\end{figure}

The analysis done so far can be generalized to the case of a $d$--dimensional system made of two--body interacting particles with an 
external linear time--dependent potential (see Appendix C).

\section{Lieb--Liniger model in a time--periodic linear potential}

Let's now consider the integrable Lieb--Liniger model in a time--periodic linear potential $V(x,t)=xf(t)$, where 
we are going to show the integrability  of the corresponding Floquet Hamiltonian. The  Lagrangian density in second quantization reads 
\be
\mathcal{L} = \frac{i \hbar}{2} \left[ \psi^{\dagger}\frac{\p \psi}{\p t} - 
h.c.\right]- \frac{\hbar^2}{2m}
\frac{\p \psi^{\dagger}}{\p x}\frac{\p \psi }{\p x}
\label{Lag_LL}- \frac{\lambda}{2} \psi^{\dagger} \psi^{\dagger} \psi \psi -V(x, t)\psi^{\dagger} \psi \, ,
\ee
where $\psi (x, t)$ and $\psi^{\dagger} (x, t)$ are the bosonic annihilation and creation operators, while $h.c.$ denotes the hermitian conjugate of the first term. 
Proceeding as before, 
we can solve the Schr\"odinger equation of the many-body interacting LL model. First of all, performing the transformation (\ref{wavefunction_gaugetrasf}) for the fields
\be
\label{field_gaugetrasf}
\psi (x,t) = e^{i \theta(x,t)} \varphi(y(t), t)\,,
\ee
with $y(t) = x - \xi(t)$,  we can rewrite the Lagrangian density (\ref{Lag_LL}) in terms of $\varphi(y,t)$ which involves no longer the external potential 
\be
\nonumber
\mathcal{L} =\frac{i \hbar}{2}\left[ \varphi^{\dagger} \frac{\p \varphi}{\p t} - 
h.c.\right]- \frac{\hbar^2}{2m} \frac{\p \varphi^{\dagger}}{\p y}\frac{\p \varphi}{\p y} - \frac{\lambda}{2} \varphi^{\dagger} \varphi^{\dagger} \varphi \varphi  \,,
\label{Lag_LL_free}
\ee
provided that, for $\int_0^T f(\tau)\,d\tau=0$, the functions $\xi(t)$ and $\theta(x_j,t)$ are 
given by Eqs. (\ref{conditions_integrab_app}) in Appendix A. If $\int_0^T f(\tau)\,d\tau \neq 0$,  
at stroboscopic times $\theta$ depends also on $x$ but nevertheless one can always eliminate the resulting linear term.
Notice that this procedure also works for a more general potential of the 
form $V(x,t) = xf(t)+ g(t)$. 

A multiparticle state $\ket{\psi}$ of the periodically driven LL model can be written  as 
\be
\nonumber
\ket{\psi} = \frac{1}{\sqrt{N!}} \int dy_1 \dots dy_N \,\eta(y_1,\dots,y_N,t)\, \varphi^{\dagger}(y_1,t)\dots\varphi^{\dagger}(y_N,t)\ket{0}\,,
\ee
where 
\be
\label{multipart_states}
\chi(x_1,\dots,x_N,t)\equiv \prod_{i=1}^{N} e^{i \theta(x_i ,t)} \eta(y_1,\dots,y_N,t)\,,
\ee
with $\chi$ solution of the Schr\"odinger equation 
$i\hbar\frac{\p}{\p t} \chi(x_1,\dots,x_N,t)=H \chi(x_1,\dots,x_N,t)$, 
with Hamiltonian $H=-\frac{\hbar^2}{2m} \sum_{j=1}^{N} \frac{\p^2}{\p x_j^2} + \lambda \sum_{j<i} \delta(x_j -x_i) + \sum_{j=1}^N V(x_j,t)$
while $\eta$ is the solution of the 
Schr\"odinger equation with no external potential ($V\,=\,0$).

\section{Floquet Hamiltonian}

The Floquet Hamiltonian ruling the stroboscopic time evolution of the many--body wavefunction 
is then  
\be
\label{Floq_HamG_manybody}
H_F=\sum_{j=1}^N \left( \frac{\hat{p}_j^2}{2\,m}+\hat{p}_j \frac{\xi(T)}{T}-
\hbar \frac{ \theta(T)}{T}\right)+\lambda\sum_{j<i} \delta(x_j-x_i)\,.
\ee
This expression for $H_F$ can be written in terms of the effective mass 
$m_{\rm eff}$ entering Eq. (\ref{m_eff_1body}).

We can now apply the standard Bethe ansatz technique (see \cite{Korepin1993}) to compute the quasi-energies: for instance, when 
$f(t)=\ell \sin(\omega t)$ their expression is 
\be
\label{quasien_manybody}
\mathcal{E}_F=\frac{\hbar^2}{2m}\sum_{j=1}^N k_j^2-\frac{\ell\,\hbar}{m\,\omega}\sum_{j=1}^N k_j+\frac{3\,\ell^2 N}{4m\,\omega^2}\,. 
\ee
If the system is subjected to periodic boundary conditions (PBC), the pseudo-momenta $k_j$ are determined in terms of the Bethe equations 
 \cite{LiebLiniger1963}, i.e. they are solutions of the transcendental equation
\be
\label{Bethe_eqs}
k_j\,L+2\sum_{i=1}^N {\rm \arctan}\left[\frac{2\,\hbar^2(k_j-k_i)}{m\,\lambda}\right]=2\pi\left(j-\frac{N+1}{2}\right) \,,
\ee
for $j=1,\dots,N$, where $L$ is the circumference of the ring in which the system is confined. The total momentum is given by $\langle \hat{P} \rangle=\hbar \sum_{j=1}^N k_j$ 
and one can see that when $\int_0^T f(\tau)\,d\tau=0$, the expectation value of the momentum doesn't change at the stroboscopic times. The center of mass in the stroboscopic dynamics is moving with constant velocity, and the width and other quantities are given by the values obtained for $V=0$. 

\begin{figure*}[t]
\centering
\includegraphics[width = 320pt]{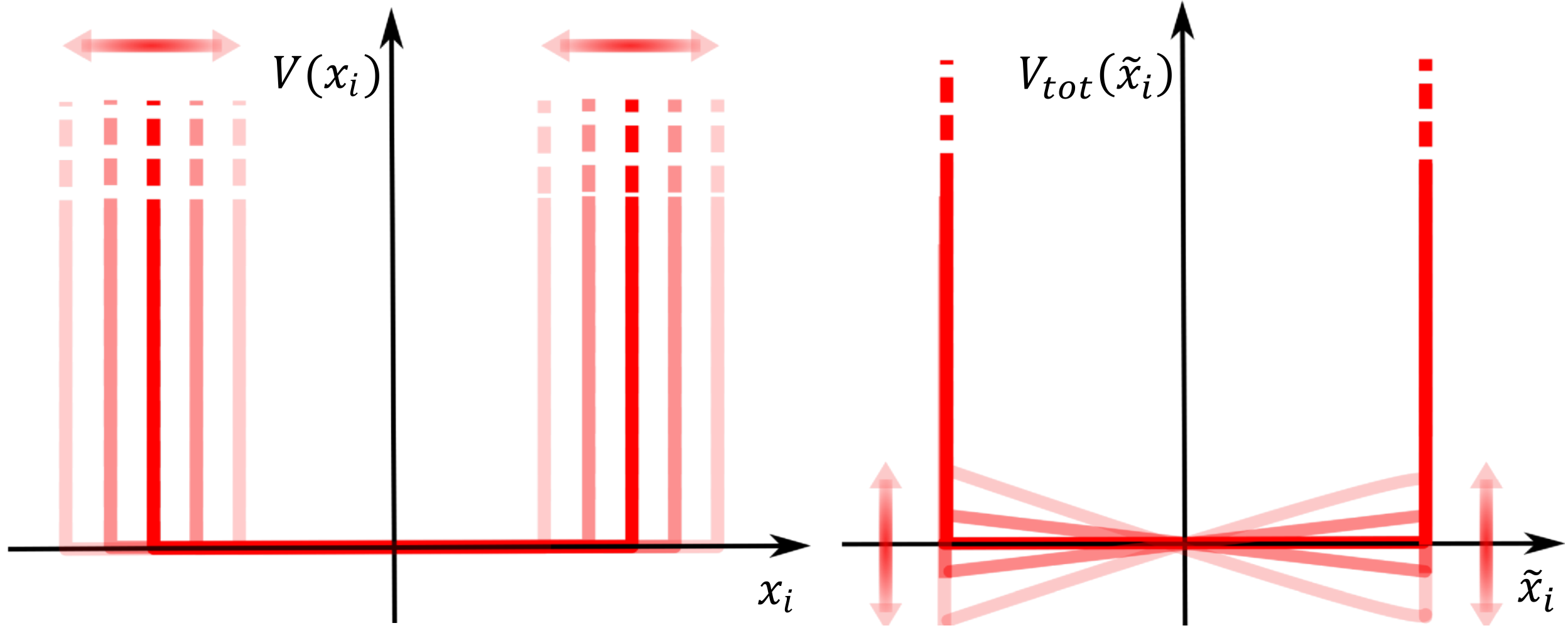}
\includegraphics[width = 160pt]{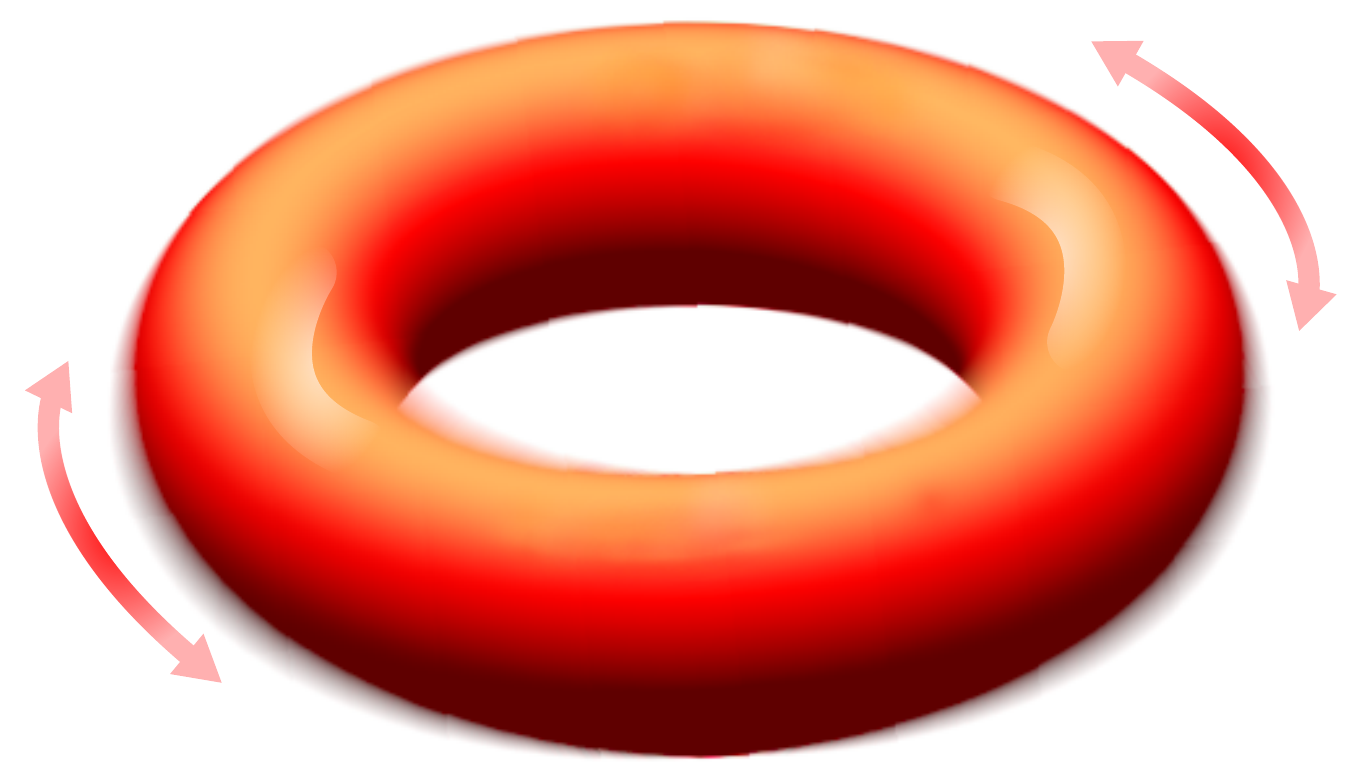}
\caption{Left: pictorial visualization of the motion of a hard wall box.
Center: the respective potential seen in the comoving frame of reference. 
Right: periodic rotation of a ring potential.}
\label{fig3}
\end{figure*} 

{\it Boundary conditions and a possible experimental realization}. An acute reader may object to the use of PBC [implemented by Eq. (\ref{Bethe_eqs})] in presence of an oscillating linear potential and, of course, he/she would have a good reason for that. Notice that confining the $1D$ Bose gas in a hard-wall potential and periodically moving the hard-wall potential back and forth [see Fig. \ref{fig3}-left, center], the proper boundary conditions to impose on such a system are instead the open boundary conditions (OBC) (see the discussion in Appendix D). However, confining the gas in a $1D$ ring and periodically rotating the ring potential [see Fig. \ref{fig3}-right], with the condition that at stroboscopic times $f(t)$ vanishes, the proper boundary conditions of the system are the PBC. This suggests that a suitable way to experimentally realize a periodically tilted $1D$ LL model 
is to shake the confining waveguide (possibly ring-shaped to implement PBC for the LL models) in which the atoms are trapped.

\section{Conclusions}

In this paper we have studied the effect of a time--periodic tilting in the Lieb--Liniger model with repulsive interactions and we have shown that the corresponding Floquet Hamiltonian 
is integrable. We discussed the spectrum of the quasi-energies and the dynamics of the system at stroboscopic times. We have shown that the Floquet Hamiltonian 
can be written in terms of a  momentum dependent effective mass. Moreover, we have also proposed an experimental setting to implement a time--periodic 
tilting using ultracold atoms in confined in $1D$ traps.
We finally observe that the analysis presented here for the Lieb--Liniger model can be 
extended to other $1D$ integrable models in time--periodic linear potentials 
as, for instance, the Gaudin-Yang model for fermions. 
 
\acknowledgements
Discussions with J. Schmiedmayer and M. Aidelsburger are gratefully acknowledged. The authors thank the Erwin Schr\"odinger Institute (ESI) in Wien for support during the Programme ``Quantum Paths''. GS acknowledges financial support from the grants FIS2015-69167-C2-1-P, 
and SEV-2016-0597 of the "Centro de Excelencia Severo Ochoa"  Programme.

\section*{Appendix A}

\setcounter{equation}{0}

\renewcommand{\theequation}{A.\arabic{equation}}

The Schr\"odinger equation for a particle of mass $m$ in 
a linear time--periodic external potential in $1D$ is given by 
\be
\label{onebody_schro_app}
i \hbar \frac{\p \chi}{\p t}= -\frac{\hbar^2}{2\,m} \frac{\p^2 \chi}{\p x^2} + x\,f(t)\,\chi(x,t)\,,
\ee
where $f(t)=f(t+T)$ is a time periodic function of period $T$. 
We put
\be
\label{wavefunction_gaugetrasf_app}
\chi(x,t) = e^{i \theta(x,t)} \,\eta(y(t), t)\,,
\ee
with $y(t)=x-\xi(t)$ and $\xi(t)$ and $\theta(x,t)$ functions to 
be determined later. Then substituting (\ref{wavefunction_gaugetrasf_app}) 
into (\ref{onebody_schro_app}), we find that $\eta(y,t)$ 
satisfies the Schr\"odinger equation with no external potential in the 
new spatial variables:
\be
\label{onebody_schro_eta_app}
i \hbar \frac{\p \eta}{\p t}= -\frac{\hbar^2}{2\,m} \frac{\p^2 \eta}{\p y^2}\,,
\ee
if we impose
\be
\label{conditions_integrab_app}
\frac{d\xi}{dt} =  \frac{\hbar}{m} \frac{\p \theta}{\p x} \,, \,\,\,\,\,\,\,\,\,\,\,\,\, -\hbar \frac{\p \theta}{\p t} = \frac{\hbar^2}{2m} \left(\frac{\p \theta}{\p x}\right)^2 + x\,f(t)\,.
\ee
Making the ansatz 
\be
\label{ansatz_theta_app}
\theta(x,t) = \frac{m}{\hbar} \frac{d\xi}{dt} \,x+ \Gamma(t)\,,
\ee
one finds the conditions
\be
\label{vANDdelta_onebody_app}
m\frac{d^2 \xi}{dt^2} = -f(t)\,, \,\,\,\,\,\,\,\,\,\, \hbar \frac{d \Gamma}{dt}=-\frac{m}{2} \left(\frac{d\xi}{dt} \right)^2 \, , 
\ee
which determine the functions $\xi(t)$ and $\Gamma(t)$ in terms of $f(t)$. 
Solving the differential equations (\ref{vANDdelta_onebody_app}) 
with the initial conditions 
$\xi(0)\,= d\xi(0)/dt= 0$ and $\Gamma(0)=0$, we get 
the following expression for the gauge phase
\be
\label{theta_f(t)_app}
\theta(x,t)=-\frac{x}{\hbar} \int_0^t f(\tau)\,d\tau -\frac{1}{2\,m\,\hbar} \int_0^t \left[\int_0^\tau f(\tau')\,d\tau' \right]^2\,d\tau\,,
\ee
which, together with (\ref{wavefunction_gaugetrasf_app}) 
and (\ref{onebody_schro_eta_app}), completely solves (\ref{onebody_schro_app}) 
since $\eta(y,t)$ is simply the time-dependent solution of the free 
Schr\"odinger equation. Notice that, with our choices, $\theta(x,0)=0$ and 
$y(0)=x$, therefore from (\ref{wavefunction_gaugetrasf_app}) we have that
\be
\label{initial_conditions_wf_app}
\chi(x,0)=\eta(x,0)\,.
\ee

Since (\ref{onebody_schro_app}) is a differential equation with a periodic 
coefficient, we may employ the Floquet approach, 
according to which the Floquet Hamiltonian, $H_F$, 
controls the time evolution of the wavefunction at stroboscopic times 
$t=n T$, with $n\in \mathbb{N}$, as
\be
\label{Floq_Ham_def_app}
\chi(x,nT)=e^{-i\frac{nT}{\hbar}H_F}\,\chi(x,t=0)\,.
\ee
The eigenvalues of the Floquet Hamiltonian 
will be denoted by $\mathcal{E}_F$. They are the time-like analogues of the quasi-momenta in the study of crystalline solids.

Let's now assume 
\be
\label{condition_f(t)_app}
\int_0^T f(t) dt = 0\,:
\ee
The gauge phase (\ref{theta_f(t)_app}) 
does not depend anymore on the spatial variable, 
{\it i.e.} $\theta(x,nT)\equiv\theta(nT)$. One has
\be
\label{theta_final_app}
\theta(nT) =-\frac{1}{2m\,\hbar}\int_0^{nT} dt \left[\int_0^t dt'f(t') \right]^2\,,
\ee
and 
\be
\label{v_final_app}
\xi(nT)=-\frac{1}{m}\int_0^{nT} dt \int_0^t dt'f(t')\,,
\ee
therefore from (\ref{Floq_Ham_def_app}) 
we have the following identification for the Floquet Hamiltonian
\be
\label{Floq_HamG_onebody_app}
H_F=\frac{\hat{p}^2}{2m}+\hat{p}\,\frac{\xi(nT)}{nT}-\hbar\frac{\theta(nT)}{nT}\,.
\ee
To see that the ratios $\frac{\xi(nT)}{nT}$ and $\frac{\theta(nT)}{nT}$ 
do not depend on time, {\it i.e.} on $n$, 
let us define a function $\mathcal{F}(t)$ such that 
$\frac{d\mathcal{F}}{dt}=f(t)$. The constant of integration is chosen to be 
such that $\mathcal{F}(0)=0$. From (\ref{condition_f(t)_app}) one has
\be
\int_0^T f(t)dt=\mathcal{F}(T)=0\,,
\ee
which implies that $\mathcal{F}(t)$ is a periodic function of period $T$. 
Using the definition of the function $\mathcal{F}$, from (\ref{v_final_app}) it follows that
\be
\xi(nT)=-\frac{1}{m}\,\int_0^{nT}\mathcal{F}(t)dt=-\frac{I}{m}n\,,
\ee
where we defined the constant $I\equiv\int_0^T\,\mathcal{F}(t)dt$. 
Hence the ratio $\frac{\xi(nT)}{nT}$ is $n$--independent. 
The same reasoning applies for the gauge phase, where we obtain
\be
\theta(nT)=-\frac{I'}{2\,m\,\hbar}\,n\,,
\ee
where $I' \equiv\int_0^T\mathcal{F}^2(t)dt$. 
Since neither $\frac{\xi(nT)}{nT}$, nor $\frac{\theta(nT)}{nT}$ 
depends on $n$, the Floquet Hamiltonian (\ref{Floq_HamG_onebody_app}) 
does not vary with time.

\begin{figure}[t]
\includegraphics[width=280pt]{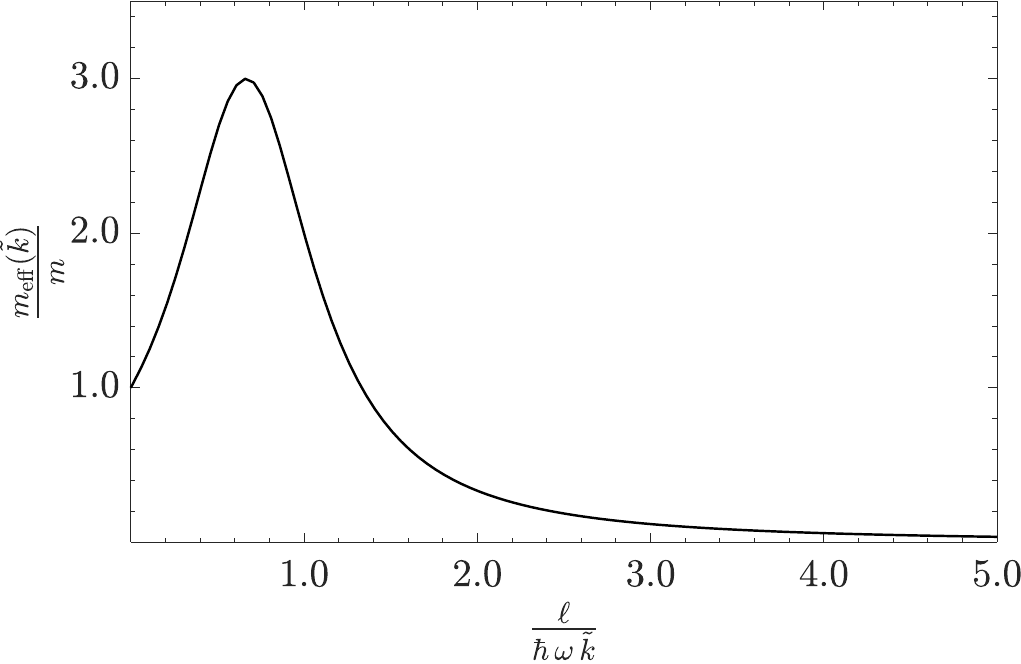}
\caption{Ratio between the effective mass of a particle observed at stroboscopic times and the physical mass {\it vs} 
the dimensionless ratio $\frac{\ell}{\hbar\, \omega\,\tilde{k}}$. 
The curve reaches its maximum at $\frac{\ell}{\hbar\, 
\omega\,\tilde{k}}=\frac{2}{3}$ for which the function is equal to $3$, and remains larger than $1$ for $\frac{\ell}{\hbar\, \omega\,\tilde{k}} < \frac{4}{3}$.}
\label{fig4}
\end{figure}

As an example we study the case in which 
\be
\label{f(t)_sin_app}
f(t)=\ell\,\sin(\omega t)\,,
\ee
with $\omega=\frac{2\pi}{T}$ and the parameter $\ell$ has the dimension of energy divided by length. Substituting (\ref{f(t)_sin_app}) in the expression 
for $m_{\rm eff}(\tilde{k})$ given in the main text, one obtains 
\be
\label{m_eff/m_onebody_f(t)_app}
\frac{m_{\rm eff}(\tilde{k})}{m} =\left[1-\frac{2\,\ell}{\hbar\,\omega\,\tilde{k}}+\frac{3}{2}\left(\frac{\ell}{\hbar\,\omega\,\tilde{k}}\right)^2 \right]^{-1}\,,
\ee
plotted in Fig. \ref{fig4}.

Notice that for $\frac{\ell}{\hbar\,\omega\,\tilde{k}} < \frac{4}{3}$ and positive, the effect of the driving at stroboscopic times can be recast in a Hamiltonian of a single non-oscillating particle with a mass greater with respect to the physical one $m_{\rm eff}(\tilde{k})>m$. In other words, if the frequency of the oscillating potential is large enough (not infinity) the system can be described as a non-oscillating system with a smaller kinetic energy due to the fast driving. 
The fact that for $\omega \rightarrow \infty$ one gets an equivalence 
between $m$ and $m_{\rm eff}$ can be proved as follows: 
From (\ref{Floq_Ham_def_app}), for short $T$ one can truncate 
the Magnus expansion \cite{Blanes2009} at the lowest level in order to write 
the time--evolution operator as
\be
U(T,0)\,\approx\, e^{-\frac{i}{\hbar} \int_0^T H(\tau) \,d\tau}\,,
\ee 
which implies
\be
H_F \, \approx \, \frac{1}{T} \int_0^T H(\tau)\,d\tau\,.
\ee
Inserting in the above equation the Hamiltonian of (\ref{onebody_schro_app}) with $f(t)$ given by (\ref{f(t)_sin_app}), we obtain
\be
H_F=\frac{\hat{p}^2}{2\,m}\,,
\ee
therefore $m=m_{\rm eff}(\tilde{k})$ if we look at the system after one short period of oscillation.

\section*{Appendix B}
\setcounter{equation}{0}

\renewcommand{\theequation}{B.\arabic{equation}}
Let's consider a system of two--body interacting particles 
in oscillating time--periodic potential, described by the following 
Schr\"odinger equation
\be
\label{Schro_twobodies_generic_app}
i\,\hbar\,\frac{\p \chi}{\p t}=\bigg\{-\frac{\hbar^2}{2\,m}\left[\frac{\p^2}{\p x_1^2}+\frac{\p^2}{\p x_2^2} \right]+V(x_1,t)+V(x_2,t) +\lambda\,\delta(x_2-x_1) \bigg\}\chi\,,
\ee
where 
\be
\label{pot_text}
V(x_j,t)=x_jf(t)\,.
\ee

To work out the dynamics starting from a general initial wavepacket, we 
introduce as usual the relative and center of mass variables $x=x_2-x_1$ and 
$X=\frac{x_2+x_1}{2}$. The Floquet Hamiltonian reads then 
\be
H_F \equiv H_{\rm com}+H_{\rm rel}\,,
\ee
where the Hamiltonian describing the center of mass of the system is simply 
\be
\label{H_com}
H_{\rm com}=-\frac{\hbar^2}{4m}\frac{\p^2}{\p X^2}-i\hbar\,\frac{\xi(nT)}{nT}\frac{\p}{\p X} -2\hbar\frac{\theta(nT)}{nT}\,,
\ee
while the one about the relative motion reads
\be
\label{H_rel}
H_{\rm rel}= -\frac{\hbar^2}{m}\frac{\p^2}{\p x^2}+\lambda \delta(x)\,.
\ee
In this way we are able to write and study separately 
the time evolution of the wavefunction at stroboscopic times as
\be
\chi(x_1,x_2,nT)=e^{-i\frac{nT}{\hbar}H_F}\,\chi(x_1, x_2,t=0) \equiv e^{-i\frac{nT}{\hbar}H_{\rm com}}\Phi(X,0) e^{-i\frac{nT}{\hbar}H_{\rm rel}}\varphi(x,0)\,,
\ee
where we denoted with $\Phi(X,t)$ the center of mass wavefunction and with $\varphi(x,t)$ the wavefunction of the relative motion.

We are going to study the stroboscopic time evolution of a Gaussian wavepacket
\be
\label{gauss_wp_app}
\chi(x_1,x_2,0)=\frac{1}{\sqrt{\pi\sigma^2}}\,e^{-\left(x_1^2 +x_2^2\right)/2\sigma^2}=\sqrt[4]{\frac{2}{\pi \sigma^2}}\,e^{-X^2/\sigma^2}\,\frac{1}{\sqrt[4]{2\pi\sigma^2 }}\,e^{-x^2/4\sigma^2}\,,
\ee
where $\sigma$ is the variance of the initial wavepacket. 
We have to describe the time evolution of each part of the wavefunction, 
{\it i.e.} relative and center of mass.
For the latter, we have
\be
\label{basis_com}
\Phi(X,nT)=\frac{1}{\sqrt{2\pi}}\,e^{-i\frac{nT}{\hbar}\,\left[\frac{\hbar^2 k^2}{4\,m}+\hbar\,k\frac{\xi(nT)}{nT}-2\hbar\frac{\theta(nT)}{nT} \right]}\,e^{i\,k\,X}\,,
\ee
where $k=\sqrt{4mE_{com}/\hbar^2}$ as usual.
Expanding the Gaussian $\Phi(X,0)=\sqrt[4]{\frac{2}{\pi \sigma^2}}\,e^{-X^2/\sigma^2}$ as usually done in Quantum Mechanics textbooks 
(see \cite{Griffith} as an example), at the end of the 
straightforward calculation, using $f(t)=\ell \sin{(\omega t)}$, we get the result
\be
\label{com_evolved}
\Phi(X,nT)=\sqrt[4]{\frac{2}{\pi \sigma^2}}\,\frac{e^{-i \frac{3 \ell^2 nT}{2 m \hbar \omega^2}}}{\sqrt{1+i\hbar \frac{nT}{m \sigma^2}}} \exp\left\{-\frac{(X+\frac{\ell\,nT}{m\omega})^2}{\sigma^2 \left(1+\frac{i\hbar\,nT}{m\sigma^2}\right)}\right\}\,.
\ee

For the relative motion instead [see Eq. (\ref{H_rel})], 
we may rely on the fact that the propagator $G(x,x';t,0)$ with 
which a generic wavefunction evolves in presence of an external Dirac 
$\delta$-potential is known in literature 
\cite{Bauch1985,Andreata2004}. It is 
\be
\label{propagator_eq}
\varphi(x,nT)=\int_{-\infty}^{\infty} G(x,x';nT,0)\,\varphi(x',0)\,dx'\,,
\ee
where
\be
\label{propagator}
G(x,x';t,0)=\frac{1}{\sqrt{4\,\pi\,i\,\hbar\,t/m}}\,e^{i\,\frac{m\,(x-x')^2}{4\,\hbar\,t}}-\frac{m\,\lambda}{4\,\hbar^2}\,e^{\frac{m\,\lambda}{2\,\hbar^2}\left(\left|x\right| +\left|x'\right|\right) +i\,\frac{m\,\lambda^2\,t}{4\,\hbar}}\,{\rm erfc}\left(\frac{\left| x\right|+\left| x'\right|+i\frac{\lambda\,t}{\hbar}}{\sqrt{4\,i\,\hbar\,t/m}}\right)\,,
\ee
where ${\rm erfc}$ is the complementary error function
\be
\nonumber
{\rm erfc}(z)=\frac{2}{\sqrt{\pi}}\int_z^\infty e^{-t^2}\,dt\,.
\ee
Then by numerically solving the integral in 
(\ref{propagator_eq}), together with (\ref{com_evolved}), 
we know how the complete wavefunction $\chi(x_1,x_2,nT)$ 
evolves starting from (\ref{gauss_wp_app}) for every $n$. In the case of hard--core interactions, 
$\lambda=\infty$, one can work out analytically 
the integral in (\ref{propagator_eq}) and the following expression is found
\be
\label{relative_twobody_TG}
\varphi(x,nT)=\frac{1}{\left(2\pi\right)^{1/4}}\sqrt{\frac{i m \sigma/\hbar t}{-1+i m \sigma^2/\hbar t}}\, \,{\rm erf}\left(\frac{m\,\sigma\,x}{2\hbar t \sqrt{-1+i m\sigma^2/\hbar t}} \right)\,e^{-\frac{m}{4\hbar t}\frac{x^2}{i+m\sigma^2/\hbar t}}\,,
\ee
where ${\rm erf}(z)=1-{\rm erfc}(z)$.

We studied the density matrix
\be
\label{density_matrix_def}
\rho(x_1,nT)=2\,\int_{-\infty}^\infty \left| \chi(x_1,x_2,nT)\right|^2\,dx_2\,,
\ee
and the time evolution of the variance and expectation values of $x_j$, 
for $j=1,2$. We evaluated the density matrix (\ref{density_matrix_def}) 
with respect to the center of mass and relative wavefunctions as
\be
\label{density_matrix_calc}
\rho(x_1,nT)=2\,\int_{-\infty}^\infty \left| \Phi\left(\frac{x}{2}+x_1,nT\right)\right|^2\,\left|\varphi(x,nT) \right|^2\,dx\,,
\ee
and we defined a length scale $\Lambda$ with respect to which we can pass to dimensionless variables (denoted by tildas): $x=\Lambda \tilde{x}$, 
$\frac{\hbar\,T}{m}=\Lambda^2 \,\tilde{T}$, $\frac{m\lambda}{\hbar^2}=
\frac{\tilde{\lambda}}{\Lambda}$, $\sigma=\Lambda\tilde{\sigma}$, 
$\frac{m\,\ell}{\hbar^2}=\frac{\tilde{\ell}}{\Lambda^4}$. 
The results are reported in Fig. \ref{fig1} for different times and values of the coupling strength $\tilde{\lambda}$.

Moreover, defining
\be
\label{exp_value_n}
\left\langle x_j^{\mathcal{N}}\right\rangle(nT)=\int_{-\infty}^\infty dx_1 \int_{-\infty}^{\infty} dx_2\, \left| \chi(x_1,x_2,nT)\right|^2 x_j^{\mathcal{N}}\,
\ee
with $\mathcal{N}=1,2$ and $j=1,2$, and $\Delta x_j^2 (nT)= \left \langle x_j^2 \right\rangle (nT)- \left[\left\langle x_j\right\rangle(nT)\right]^2$, it is found 
\be
\label{mean_values}
\left\langle x_j\right\rangle(nT)=-\frac{\ell\,nT}{m\,\omega}\,.
\ee
The motion of the wavepacket's center is controlled 
only by the center of mass motion, {\it i.e.} $\left\langle 
x\right\rangle(nT)=0$, and therefore one can simply write 
$\left\langle x_j\right\rangle(nT)=\left\langle X\right\rangle(nT)$. 
The evolution of the variance is instead controlled both by the center of 
mass contribute, for which one has the usual free Gaussian spreading
\be
\label{variance_X}
\Delta X(nT)=\sqrt{\left\langle X^2\right\rangle(nT) -\left[\left\langle X\right\rangle(nT)\right]^2}\,=\,\frac{\sigma}{2}\,\sqrt{1+\left(\frac{\hbar\,nT}{m\,\sigma^2}\right)^2}\,,
\ee
and by the relative motion part which depends on $\lambda$. 
In particular one has
\be
\label{variance_xj}
\Delta x_j (nT)=\sqrt{\left[\Delta X(nT)\right]^2+\frac{1}{4}\left[\Delta x (nT)\right]^2}\,,
\ee
whose behaviour is reported in Fig. \ref{fig1}\circled{d} for different interactions. 

For $\lambda=0$ also the relative motion spreading follows the free 
Gaussian one, therefore one may explicitly write
\be
\label{variance_check_app}
\Delta x_j^0 (nT)=\frac{\sigma}{\sqrt{2}} \sqrt{1+\left(\frac{\hbar nT}{m \sigma^2}\right)^2}\,,
\ee
which is reported in Fig. \ref{fig1}\circled{d} in black solid line. 

\section*{Appendix C}
\setcounter{equation}{0}

\renewcommand{\theequation}{C.\arabic{equation}}
The analysis presented in the main text may be generalized 
to two--body interacting particles with an external linear time--dependent potential in $d$--dimensions, described by the following Schr\"odinger equation 
\be
\label{Schro_twobodies_d-dim}
i \hbar \frac{\p \chi}{\p t}=\bigg\{-\frac{\hbar^2}{2\,m} \vec{\nabla}_x^{\,2}+\sum_{\alpha=1}^{d}\left[\left(x_1^{(\alpha)}+x_2^{(\alpha)}\right) f^{(\alpha)}(t)\right] +V_{2b}(\left|\vec{x}_2-\vec{x}_1\right|) \bigg\}\chi\,,
\ee
where $\alpha=1,\dots,d$ denotes the spatial dimension, $x_{i}^{(\alpha)}$ 
the coordinates of the $i$-th particle ($i=1,2$), and 
$V_{2b}(\left|\vec{x}_2-\vec{x}_1\right|)$ is a generic interacting 
potential depending on the distance between the two particles
\be
\nonumber
\left|\vec{x}_2-\vec{x}_1\right|=\sqrt{\sum_{\alpha=1}^d \left(x_1^{(\alpha)}-x_2^{(\alpha)}\right)^2} \,.
\ee
The kinetic energy term is written for the sake of brevity as
\be
\nonumber
\vec{\nabla}_x^{\,2}=\sum_{\alpha=1}^d \left[\frac{\p^2}{\p \left(x_1^{(\alpha)}\right)^2}+\frac{\p^2}{\p \left(x_2^{(\alpha)}\right)^2} \right]\,.
\ee
In (\ref{Schro_twobodies_d-dim}) the external potential could 
have a different time dependence in different directions, 
due to the indexing of the functions $f(t)$, which could be for example
\be
f^{(\alpha)}(t)=\ell^{(\alpha)} \sin(\omega^{(\alpha)} t)\,,
\ee
as a generalization of (\ref{f(t)_sin}). Therefore, 
in order to generalize the gauge transformation in 
(\ref{wavefunction_gaugetrasf}), one should introduce $d$ different 
gauge phases and write the wavefunction as
\be
\chi\left(\vec{x}_1,\vec{x}_2,t\right)=e^{\sum_{\alpha=1}^d\left[\theta^{(\alpha)}\left(x_1^{(\alpha)}, t \right) +\theta^{(\alpha)}\left(x_2^{(\alpha)} , t\right)\right]} \eta(\vec{y}_1, \vec{y}_2, t)\,,
\label{wavefunction_gaugetrasf_twobodies_d-dim}
\ee
where  
\be
y_i^{(\alpha)}(t)=x_i^{(\alpha)}-\xi^{(\alpha)}(t)\,,\,\,\,\,\,\,\,\forall \alpha=1,\dots,d\,,\,\,\,i=1,2\,.
\ee
Using (\ref{wavefunction_gaugetrasf_twobodies_d-dim}) in (\ref{Schro_twobodies_d-dim}), one has that the wavefunction $\eta$ satisfies a similar Schr\"odinger equation but with no external potential, {\it i.e.}
\be
\label{Schro_twobodies_d-dim_eta}
i\hbar\frac{\p \eta}{\p t}=\bigg\{-\frac{\hbar^2}{2m} \vec{\nabla}_y^{\,2}+V_{2b}(\left|\vec{y}_2-\vec{y}_1\right|) \bigg\}\eta\,,
\ee
if the following equations are satisfied for every $\alpha=1,\dots,d$
\begin{gather}
\nonumber
\frac{d\xi^{(\alpha)}}{dt} = \frac{\hbar}{m} \frac{\p \theta^{(\alpha)}}{\p x^{(\alpha)}} \,, 
\\ -\hbar \frac{\p \theta^{(\alpha)}}{\p t} = \frac{\hbar^2}{2m} \left(\frac{\p \theta^{(\alpha)}}{\p x^{(\alpha)}}\right)^2 + x^{(\alpha)}\,f^{(\alpha)}(t)\,,
\label{theta_condition_d-dim}
\end{gather}
valid both for $x^{(\alpha)}=x_1^{(\alpha)}$ and $x_2^{(\alpha)}$. 
To get an expression for $\xi^{(\alpha)}$ and $\theta^{(\alpha)}$ 
with respect to the external potential parameters, we perform 
the following ansatz for the gauge phase 
$\forall \alpha=1,\dots,d$
\be
\theta^{(\alpha)}\left(x^{(\alpha)},t\right) = \frac{m}{\hbar} \frac{d\xi^{(\alpha)}}{dt} \,x^{(\alpha)}+ \Gamma^{(\alpha)}(t)\,.
\ee
We then obtain a set of differential equations 
from (\ref{theta_condition_d-dim})
\be
m\frac{d^2 \xi^{(\alpha)}}{dt^2} = -f^{(\alpha)}(t)\,, \,\,\,\,\,\,\,\,\,\, \hbar \frac{d \Gamma^{(\alpha)}}{dt}=-\frac{m}{2} \left(\frac{d\xi^{(\alpha)}}{dt} \right)^2 \, ,
\ee
decoupled in terms of the index $\alpha$. 
Solving these differential equations with the initial conditions 
$\xi^{(\alpha)}(0)\,= d\xi^{(\alpha)}/dt(0) =  0$ and $\Gamma^{(\alpha)}(0)=0$ $\forall \alpha$, one obtains in the different directions
\be
\theta^{(\alpha)}\left(x^{(\alpha)},t\right)=-\frac{x^{(\alpha)}}{\hbar} \int_0^t f^{(\alpha)}(\tau)\,d\tau -\frac{1}{2m\hbar} \int_0^t \left[\int_0^\tau f^{(\alpha)}(\tau')\,d\tau' \right]^2,\,d\tau\,,
\ee
for $x^{(\alpha)}=x_1^{(\alpha)}$ and $x_2^{(\alpha)}$. 
If one is able to solve (\ref{Schro_twobodies_d-dim_eta}), 
using the above expression for $\theta^{(\alpha)}$'s 
together with (\ref{wavefunction_gaugetrasf_twobodies_d-dim}), 
then a solution for the Schr\"odinger equation (\ref{Schro_twobodies_d-dim}) 
can be written.

\section*{Appendix D}
\setcounter{equation}{0}

\renewcommand{\theequation}{D.\arabic{equation}}

We discuss how one could obtain 
periodic boundary conditions (PBC) in a system subjected to a 
time--dependent linear external potential. Before doing that, 
is better to start the discussion with open boundary conditions (OBC).

Let us first consider the case of a Schr\"odinger equation describing a moving hard wall box with one--particle inside
\be
\label{1body_movingHWB}
i\hbar\frac{\p}{\p t} \chi(x,t)=\left[-\frac{\hbar^2}{2m} \frac{\p^2}{\p x^2} +V\left(x,t\right) \right]\chi(x,t)\,,
\ee
where
\be
\label{moving_hardwallbox}
V(x,t)=
\begin{cases}
	0 & \text{for $-\frac{L}{2} +h(t)<x< \frac{L}{2}+h(t)$}\\
	\infty & \text{elsewhere}
\end{cases}
\,.
\ee
The function $h(t)$ describes the motion of the box. For example in the case 
in which the box translates with constant acceleration $a$, 
then $h(t) \propto a\frac{t^2}{2}$, while if the box oscillates 
around the origin with a maximum amplitude of oscillation $x_0$ and 
frequency $\omega$, then $h(t)=x_0\,\sin(\omega\,t)$. 
Because of the potential (\ref{moving_hardwallbox}), the wavefunction will satisfy the following moving OBC
\be
\label{OBC_movingHWB}
\chi(x=-L/2+h(t),t)=\chi(x=L/2+h(t),t)\,=\,0\,.
\ee
Let's describe the system in the comoving frame of reference. This can be done by changing the spatial variable from $x$ to 
\be
\label{comoving_OBC}
\tilde{x}=x-h(t)\,,
\ee
thanks to which we pass from (\ref{1body_movingHWB}) to 
\be
\label{comoving_1body_HWB}
i\hbar\frac{\p}{\p t} \chi(\tilde{x},t)=\left[-\frac{\hbar^2}{2m} \frac{\p^2}{\p \tilde{x}^2} +V(\tilde{x}) +i\hbar\,\frac{dh}{dt}\,\frac{\p}{\p \tilde{x}} \right]\chi(\tilde{x},t)\,,
\ee
and now the wavefunction satisfies the fixed boundary conditions
\be
\chi(\tilde{x}=-L/2,t)=\chi(\tilde{x}=L/2,t)\,=\,0\,.
\ee
To get rid of the term $\propto\hat{p}$ in (\ref{comoving_1body_HWB}), we transform the wavefunction as
\be
\label{first_transform}
\chi(\tilde{x},t) \equiv e^{i\beta(t)\tilde{x}}\,\tilde{\chi}(\tilde{x},t)\,,
\ee
where $\beta(t)$ is a function to be determined in order to eliminate the term linear in the momentum. 
Substituting (\ref{first_transform}) in the Schr\"odinger equation (\ref{comoving_1body_HWB}) and choosing $\beta(t)\,=\,\frac{m}{\hbar} \frac{dh}{dt}$, we get
\be
i\hbar \frac{\p}{\p t} \tilde{\chi}(\tilde{x},t) = \left[-\frac{\hbar^2}{2m} \frac{\p^2}{\p \tilde{x}^2}+V(\tilde{x})+m \frac{d^2 h}{dt^2}\, \tilde{x} -\frac{3}{2}m \left(\frac{dh}{dt}\right)^2 \right]\tilde{\chi}(\tilde{x},t)   \,.
\ee
If we now perform the following transformation of the wavefunction
\be
\tilde{\chi}(\tilde{x},t)=e^{\frac{i}{\hbar}\frac{3m}{2}\int_0^t \left(\frac{dh}{d\tau}\right)^2\,d\tau}\,\bar{\chi}(\tilde{x},t)\,,
\ee
we arrive at the Schr\"odinger equation describing a particle enclosed in a fixed hard wall box and subjected to the action of a linear external potential, which reads
\be
i\hbar \frac{\p}{\p t} \bar{\chi}(\tilde{x},t) =
\\ \left[-\frac{\hbar^2}{2m} \frac{\p^2}{\p \tilde{x}^2}+V(\tilde{x})+m \frac{d^2 h}{dt^2}\, \tilde{x} \right]\bar{\chi}(\tilde{x},t)   \,,
\ee
and the wavefunction fulfills
\be
\bar{\chi}(\tilde{x}=-L/2,t)=\bar{\chi}(\tilde{x}=L/2,t)\,=\,0\,.
\ee
The presence of the external potential linear in $\tilde{x}$ is coming from the fact that passing to the comoving frame of reference [via the spatial transformation (\ref{comoving_OBC})], the particle feels the presence of an inertial force related to the second time derivative of $h(t)$. 

The procedure described above can be extended for the $1D$ LL model of $N$ atoms in a moving box of length $L$, 
where the many--body Schr\"odinger equation reads 
\be
\label{schro_chi}
i\hbar\frac{\p}{\p t} \chi(x_1,\dots,x_N,t)=H \chi(x_1,\dots,x_N,t)\,,
\ee
with Hamiltonian
\be
\label{ham_LL_pot}
H=-\frac{\hbar^2}{2m} \sum_{j=1}^{N} \frac{\p^2}{\p x_j^2} + \lambda \sum_{j<l} \delta(x_j -x_l) + \sum_{j=1}^N V(x_j,t)\,.
\ee
where the external potential is the same as in (\ref{moving_hardwallbox}) for $x=x_j$, and the wavefunction satisfies moving OBC
\be
\label{OBC_initial}
\chi(x_1=\pm L/2 +h(t),x_2,\dots,x_N,t)=\cdots=0\,.
\ee
Passing to the comoving frame by changing the spatial variables as in (\ref{comoving_OBC}), we get
\be
\label{schroeq_N_HWB}
i\hbar \frac{\p }{\p t}\chi=\bigg\{-\frac{\hbar^2}{2m} \sum_{j=1}^N \frac{\p^2}{\p \tilde{x}_j^2} +\lambda \sum_{j<l} \delta(\tilde{x}_j -\tilde{x}_l) +\sum_{j=1}^N \left[V(\tilde{x}_j) +i\hbar \frac{d h}{d t}\frac{\p}{\p \tilde{x}_j}\right] \bigg\} \chi\,,
\ee
and the OBC gets modified into
\be
\label{OBC_moving}
\chi(\tilde{x}_1=\pm L/2,\tilde{x}_2,\dots,\tilde{x}_N,t)=\cdots=0\,.
\ee
We then transform the wavefunction as
\be
\chi(\tilde{x}_1,\dots,\tilde{x}_N,t)\equiv e^{i\frac{m}{\hbar}\frac{dh}{dt} \sum_{j=1}^N \tilde{x}_j}\,\tilde{\chi}(\tilde{x}_1,\dots,\tilde{x}_N,t)\,,
\ee
under which the Schr\"odinger equation (\ref{schroeq_N_HWB}) becomes
\be
i\hbar \frac{\p }{\p t}\tilde{\chi}=\bigg\{-\frac{\hbar^2}{2m} \sum_{j=1}^N \frac{\p^2}{\p \tilde{x}_j^2} +\lambda \sum_{j<l} \delta(\tilde{x}_j -\tilde{x}_l) +\sum_{j=1}^N \left[V(\tilde{x}_j) +m \frac{d^2 h}{d t^2} \,\tilde{x}_j \right]-\frac{3}{2}\,m\,N\left(\frac{dh}{dt}\right)^2 \bigg\} \tilde{\chi}\,.
\ee
Finally, by performing the transformation
\be
\tilde{\chi}(\tilde{x}_1,\dots,\tilde{x}_N,t)=e^{\frac{i}{\hbar}\frac{3Nm}{2}\int_0^t \left(\frac{dh}{d\tau}\right)^2\,d\tau}\,\bar{\chi}(\tilde{x}_1,\dots,\tilde{x}_N,t)\,,
\ee
we have the Schr\"odinger equation 
\be
i\,\hbar \,\frac{\p }{\p t}\bar{\chi}=\bigg\{-\frac{\hbar^2}{2\,m} \sum_{j=1}^N \frac{\p^2}{\p \tilde{x}_j^2} +\lambda \sum_{j<l} \delta(\tilde{x}_j -\tilde{x}_l) +\sum_{j=1}^N \left[V(\tilde{x}_j) +V_{fict}(\tilde{x}_j)\right] \bigg\} \,\bar{\chi}\,,
\ee
with the external linear "fictitious" potential 
\be
\label{OBC_fict_pot}
V_{fict}(\tilde{x}_j)=m\,\frac{d^2 h}{d t^2} \,\tilde{x}_j\,,
\ee
and the boundary conditions
\be
\bar{\chi}(\tilde{x}_1=\pm L/2,\tilde{x}_2,\dots,\tilde{x}_N,t)=\cdots=0\,,
\ee
{\it i.e.} fixed OBC. Therefore we can describe the LL model in 
a shaken hard wall box as a fixed one subjected to the 
action of a linear potential in the comoving frame. 
See Fig. \ref{fig3} left and center for a visualization of the case 
in which $h(t)=x_0 \,\sin(\omega\,t)$. In this case, (\ref{OBC_fict_pot}) 
is a linear time--periodic potential
\be
V_{fict}(\tilde{x}_j)=-m\,\omega^2\,x_0\,\tilde{x}_j\,\sin(\omega\,t)\,.
\ee

We now discuss the case of PBC starting from a single particle enclosed in a rotating ring of circumference $L$
\be
i\hbar \frac{\p}{\p t} \chi(\varphi,t) = -\frac{\hbar^2}{2mL^2} \frac{\p^2}{\p \varphi^2} \chi(\varphi,t)\,,
\label{schro_eq_PBC_1body}
\ee
where $\varphi$ is the angle variable such that in $x$--$y$ coordinates one would have
\be
\label{xy_varphi}
\begin{cases}
	x_i=\frac{L}{2\pi}\,\cos(\varphi_i)\\
	y_i=\frac{L}{2\pi}\,\sin(\varphi_i)
\end{cases}
\ee 
and the wavefunction satisfies moving PBC
\be
\chi\left(\varphi=0+\int_0^t \bar{\mathcal{F}}(\tau)\,d\tau,t\right)=\chi\left(\varphi=2\pi+\int_0^t \bar{\mathcal{F}}(\tau)\,d\tau,t\right)\,,
\ee
in which $\bar{\mathcal{F}}(t)$ represents the angular velocity under which the ring is rotating, see Fig. \ref{fig3} right, and will coincide with the primitive of the driving $f(t)$. Analogously to what we've done for the hard wall problem, we pass to the comoving frame of reference by changing the angle variables 
\be
\label{PBC_change_variables}
\tilde{\varphi} = \varphi -\int_0^t \bar{\mathcal{F}}(\tau)\,d\tau\,, 
\ee
under which the Schr\"odinger equation (\ref{schro_eq_PBC_1body}) becomes
\be
i\hbar \frac{\p}{\p t} \chi(\tilde{\varphi},t) = \left[-\frac{\hbar^2}{2mL^2} \frac{\p^2}{\p \tilde{\varphi}^2} +i\,\hbar\,\bar{\mathcal{F}}(t)\frac{\p}{\p \tilde{\varphi}} \right] \chi(\tilde{\varphi},t)\,,
\ee
and now we have fixed PBC
\be
\label{fixed_PBC_app}
\chi(\tilde{\varphi}=0,t)=\chi(\tilde{\varphi}=2\pi,t)\,.
\ee
Next we transform the wavefunction 
\be
\label{transform_PBC}
\chi(\tilde{\varphi},t)\equiv e^{i \frac{mL^2}{\hbar}\,\bar{\mathcal{F}}(t)\tilde{\varphi}} \, \tilde{\chi}(\tilde{\varphi},t)\,,
\ee
in such a way that the Schr\"odinger equation has a $\tilde{\varphi}$--linear dependent term
\be
\label{schro_eq_PBC_1body_intermediate}
i\hbar \frac{\p}{\p t} \tilde{\chi}(\tilde{\varphi},t) = \left[-\frac{\hbar^2}{2mL^2} \frac{\p^2}{\p \tilde{\varphi}^2} +m L^2 \frac{d\bar{\mathcal{F}}}{dt} \tilde{\varphi} -\frac{3}{2}mL^2 \bar{\mathcal{F}}^2(t) \right] \tilde{\chi}(\tilde{\varphi},t)\,.
\ee
Notice that using transformation (\ref{transform_PBC}) in the fixed PBC (\ref{fixed_PBC_app}), implies that the wavefunction $\tilde{\chi}$ satisfies twisted boundary conditions 
\be
\tilde{\chi}(\tilde{\varphi}=0,t)=e^{i\frac{2\pi mL}{\hbar}\,\bar{\mathcal{F}}(t)}\,\tilde{\chi}(\tilde{\varphi}=2\pi,t)\,.
\ee
Finally we perform the transformation on the wavefunction
\be
\tilde{\chi}(\tilde{\varphi},t)\equiv e^{\frac{i}{\hbar} \frac{3mL^2}{2}\int_0^t \bar{\mathcal{F}}^2(\tau) \,d\tau} \, \bar{\chi}(\tilde{\varphi},t)\,,
\ee
thanks to which the Schr\"odinger equation reads
\be
\label{schro_eq_PBC_1body_final}
i\hbar \frac{\p}{\p t} \bar{\chi}(\tilde{\varphi},t) = \left[-\frac{\hbar^2}{2mL^2} \frac{\p^2}{\p \tilde{\varphi}^2} +mL^2 \frac{d\bar{\mathcal{F}}}{dt} \tilde{\varphi} \right] \bar{\chi}(\tilde{\varphi},t)\,,
\ee
while the boundary conditions remain the same for the new wavefunction
\be
\label{PBC_final_BC}
\bar{\chi}(\tilde{\varphi}=0,t)=e^{i\frac{2\pi m L}{\hbar}\,\bar{\mathcal{F}}(t)}\,\bar{\chi}(\tilde{\varphi}=2\pi,t)\,.
\ee
Let's now compare with the equations of the main text (\ref{Bethe_eqs}): There we studied the case of periodic driving, \textit{i.e.} periodic angular velocity $\bar{\mathcal{F}}(t)$, imposing PBC and we have the identification $f(t) = m\frac{d\bar{\mathcal{F}}}{dt}$. Notice from (\ref{PBC_final_BC}) that in order to have PBC in the wavefunction $\bar{\chi}$ at stroboscopic times, we need to impose $\bar{\mathcal{F}}(nT)=\frac{\hbar}{m L} \bar{n}$, where $\bar{n}$ is an integer number. Moreover at $t=0$, because $\tilde{\varphi}=\varphi$ and $\chi(\varphi,0) =\tilde{\chi}(\varphi,0)= \bar{\chi}(\varphi,0)$, then the Schr\"odinger equations (\ref{schro_eq_PBC_1body_final}) and (\ref{schro_eq_PBC_1body_intermediate}) should both reduce to (\ref{schro_eq_PBC_1body}), and this happens if $\bar{\mathcal{F}}(0) = 0$ and $\frac{d \bar{\mathcal{F}}}{dt}(0) = 0$, which amounts to say that at stroboscopic times the driving function $f(t)$ should vanishes, \textit{i.e.} $f(nT)=0$, as well as $\int_0^{nT} f(\tau)\,d\tau=0$, since $\bar{\mathcal{F}}(t)$ and its derivatives are periodic functions. 

Let us now generalize to the case of LL model of $N$ atoms enclosed in a ring 
of circumference $L$ which is rotating with an angular velocity $\bar{\mathcal{F}}(t)$. The following Schr\"odinger equation holds for the system
\be
\label{Schro_PBC_initial}
i\hbar \frac{\p }{\p t}\chi=\left\{-\frac{\hbar^2}{2mL^2} \sum_{j=1}^N \frac{\p^2}{\p \varphi_j^2} +\frac{\lambda}{L} \sum_{j<i} \delta(\varphi_j -\varphi_i) \right\} \chi\,,
\ee
where $\varphi_i$ is the angle variable related to $x$--$y$ coordinates via (\ref{xy_varphi}).
We have then moving PBC, written as
\be
\label{PBC_initial}
\chi\left(\varphi_1=0+\int_0^t \bar{\mathcal{F}}(\tau)\,d\tau,\varphi_2,\dots,\varphi_N,t\right)=\chi\left(\varphi_1=2\pi+\int_0^t \bar{\mathcal{F}}(\tau)\,d\tau,\varphi_2,\dots,\varphi_N,t\right)\,.
\ee
As usual, we should pass to the comoving frame by changing the variables as (\ref{PBC_change_variables}), under which the Schr\"odinger equation becomes
\be
\label{Schro_PBC_moving}
i\hbar \frac{\p }{\p t}\chi=\bigg\{-\frac{\hbar^2}{2mL^2} \sum_{j=1}^N \frac{\p^2}{\p \tilde{\varphi}_j^2} +\frac{\lambda }{L}\sum_{j<i} \delta(\tilde{\varphi}_j -\tilde{\varphi}_i)\\
 +i \,\hbar \,\bar{\mathcal{F}}(t) \sum_{j=1}^N \frac{\p}{\p \tilde{\varphi}_j}\bigg\} \chi\,,
\ee
and the PBC reads
\be
\chi(\tilde{\varphi}_1=0,\tilde{\varphi}_2,\dots,\tilde{\varphi}_N,t)=\chi(\tilde{\varphi}_1=2\pi,\tilde{\varphi}_2,\dots,\tilde{\varphi}_N,t)\,.
\ee
If we now transform the wavefunction 
\be
\chi(\tilde{\varphi}_1,\dots,\tilde{\varphi}_N,t)\equiv e^{i\frac{m L^2}{\hbar}\left[\frac{3N}{2} \int_0^t \bar{\mathcal{F}}^2(\tau)\,d\tau+\bar{\mathcal{F}}(t)\sum_{j=1}^N \tilde{\varphi}_j\right]} \,\bar{\chi}(\tilde{\varphi}_1,\dots,\tilde{\varphi}_N,t)\,,
\ee
then we arrive to a Schr\"odinger equation describing $N$ bosons with contact interaction
\be
\label{Schro_PBC_final}
i\hbar \frac{\p }{\p t}\bar{\chi}=\bigg\{-\frac{\hbar^2}{2mL^2} \sum_{j=1}^N \frac{\p^2}{\p \tilde{\varphi}_j^2} +\frac{\lambda }{L}\sum_{j<i} \delta(\tilde{\varphi}_j -\tilde{\varphi}_i)\\
 +L^2\sum_{j=1}^N V_{fict}(\tilde{\varphi}_j)\bigg\} \bar{\chi}\,,
\ee
which are subjected to a linear external time--dependent fictitious potential
\be
\label{PBC_fict_pot}
V_{fict}(\tilde{\varphi}_j)=m\frac{d\bar{\mathcal{F}}}{dt}\, \tilde{\varphi}_j\,,
\ee
and the wavefunction satisfies twisted boundary conditions
\be
\bar{\chi}(\tilde{\varphi}_1=0,\tilde{\varphi}_2,\dots,\tilde{\varphi}_N,t)=e^{i \frac{2\pi m L}{\hbar} \bar{\mathcal{F}}(t)}
 \bar{\chi}(\tilde{\varphi}_1=2\pi,\tilde{\varphi}_2,\dots,\tilde{\varphi}_N,t)\,.
\ee
Using the same reasoning of the single particle problem, we see that in order to restore PBC at stroboscopic times we have to require that $\bar{\mathcal{F}}(nT)=0$ as well as $\frac{d\bar{\mathcal{F}}}{dt}=0$, from the comparison of the Schr\"odinger equations (\ref{Schro_PBC_moving}) and (\ref{Schro_PBC_final}), with (\ref{Schro_PBC_initial}).

If we specify to the case $L\bar{\mathcal{F}}(t)=\frac{\ell}{m \omega}\left[1-\cos(\omega\,t)\right]$, we then describe a LL model enclosed in a periodically rotating ring 
as a LL model in a fixed (not moving) ring with PBC and a 
linear potential along the ring itself. 
This last model is the same one, once some subtleties are clarified 
as we are going to discuss now, as that studied in the main text.

Indeed, one may wonder if the method to solve the LL model 
in an external potential presented previously, would still be valid 
after we introduced a spatial variable (\ref{PBC_change_variables}) which now depends 
on time. We find that for the potential in (\ref{pot_text}), 
the function $\theta(x,t)$ does get modified, but actually 
only in the $x$--linear term, which do disappear anyway when 
$\int_0^T f(\tau) d\tau=0$. 

To show this, we can repeat the same steps 
used previously in the solution of the Schr\"odigner equation 
with a $x$--linear term, taking into account the fact that now also the 
spatial variable has a time dependence. First of all it is convenient 
to pass from the angle variables to spatial one via
\be
\label{comoving_PBC_revised}
L \,\tilde{\varphi}(t) 
\rightarrow \tilde{x}(t) = x-\int_0^t \bar{\mathcal{F}}(\tau)d\tau\,,
\ee
where 
\be
\label{Omega_function}
\bar{\mathcal{F}}(t) = \frac{\ell}{m \omega} \left[1-\cos(\omega\,t)\right]\,.
\ee
according to the prescription (\ref{PBC_change_variables}). 
At stroboscopic times the potential (\ref{PBC_fict_pot}) 
reduces to the one in (\ref{pot_text}) for (\ref{f(t)_sin}), 
which is the case of interest. Notice that the potential (\ref{f(t)_sin}) 
is zero at stroboscopic times, so that there are no problems for the PBC 
at these times.

We introduce the new spatial variable $\tilde{y}(t)=\tilde{x}(t)-\xi(t)$ 
accordingly:
\begin{gather}
\nonumber
\frac{\p}{\p \tilde{x}}=\frac{\p}{\p \tilde{y}}\,,
\\
\frac{\p}{\p t}=\frac{\p \tilde{y}}{\p t}\frac{\p}{\p \tilde{y}} + \frac{\p}{\p t}=-\left[\bar{\mathcal{F}}(t) +\frac{d \xi}{d t} \right]\frac{\p}{\p \tilde{y}} + \frac{\p}{\p t}\,.
\nonumber
\end{gather}
Therefore in the procedure leading from (\ref{Lag_LL}) to (\ref{Lag_LL_free}), 
we now need to change the conditions (\ref{conditions_integrab_app}) into 
\begin{gather}
\nonumber
\bar{\mathcal{F}}(t)+\frac{d\xi}{dt} = \frac{\hbar}{m} \frac{\p \theta}{\p \tilde{x}} \,, 
\\
 -\hbar \frac{\p \theta}{\p t} = \frac{\hbar^2}{2m} \left(\frac{\p \theta}{\p \tilde{x}}\right)^2 + \tilde{x}\,\ell\,\sin(\omega\,t)\,.
\label{conditions_integrab_revised}
\end{gather}
One could then make the following ansatz for the gauge phase
\be
\label{ansatz_theta_revised}
\theta(\tilde{x},t)  = \frac{m}{\hbar} \left[\bar{\mathcal{F}}(t)+\frac{d\xi(t)}{dt}\right] \tilde{x}+ \Gamma(t)\,,
\ee
under which, by replacing into (\ref{conditions_integrab_revised}), 
we find the following equations for $\xi(t)$ and $\Gamma(t)$
\begin{gather}
\nonumber
m\left[\frac{d\bar{\mathcal{F}}}{d t}+\frac{d^2 \xi(t)}{dt^2} \right]= 
-\ell \sin(\omega t)\,, \\
\hbar \frac{d \Gamma(t)}{dt}=-\frac{m}{2} \left[\bar{\mathcal{F}}(t)+\frac{d\xi(t)}{dt} 
\right]^2 \, .
\label{vANDdelta_revised}
\end{gather}
From the first of the above equations one has $\xi(t)=0$ 
by imposing $\xi(0)= d\xi/dt(0) = 0$ as initial conditions, 
while from the second equation, using (\ref{conditions_integrab_revised}), 
one has 
\be
\label{delta_revised}
\Gamma(t)=-\frac{\ell^2}{2m\hbar} \int_0^t \left[\int_0^\tau \sin(\omega\tau')\,d\tau' \right]^2d\tau\,,
\ee
having imposed $\Gamma(0)=0$. 
Notice that $\Gamma(t)$ is not changed 
[see (\ref{theta_f(t)_app}) and (\ref{f(t)_sin_app})], 
while the term linear in $\tilde{x}$ of the gauge phase is changed 
and the whole phase reads
\be
\label{theta_final_revised}
\theta(\tilde{x},t) = \frac{m}{\hbar} \bar{\mathcal{F}}(t) \tilde{x}-\frac{\ell^2}{2m\hbar} \int_0^t \left[\int_0^\tau \sin(\omega\tau')d\tau' \right]^2d\tau\,.
\ee
Therefore notice that still $\theta(\tilde{x},nT)$ doesn't depend on $\tilde{x}$ but only on time, if we require that PBC holds at stroboscopic times. 
Summarizing, when the condition (\ref{condition_f(t)}) 
is satisfied and $f(t)$ is vanishing at the stroboscopic times 
[as it happens for $f(t) \propto \sin(\omega t)$], then  
the gauge phase term is not changed and 
the final results coming from the stroboscopic analysis are unchanged as well. 
One should then stress out the fact that in obtaining our results 
for the stroboscopic dynamics we can use PBC, 
simply because we work at times multiple of the period of oscillation.


\vspace{-5mm}

\begin{thebibliography}{XX}

\bibitem{Floquet883} 
G. Floquet, 
Ann. de l'Ecole Norm. Suppl. {\bf 12}, 47 (1883).

\bibitem{Shirley65} 
J.H. Shirley, 
Phys. Rev. B {\bf 138}, 979 (1965).

\bibitem{Grifoni98} 
M. Grifoni, P. H\"anggi, 
Phys. Rep. {\bf 304}, 229 (1998).

\bibitem{Eckardt2017} 
A. Eckardt,
Rev. Mod. Phys. {\bf 89}, 011004 (2017).

\bibitem{Oka18}
T. Oka and S. Kitamura, Annu. Rev. Condens. Matter Phys. {\bf 10}, 
387 (2019).

\bibitem{Dunlap1986} 
D.H. Dunlap and V.M. Kenkre,
Phys. Rev. B {\bf 34}, 3625 (1986).

\bibitem{Creffield2003} 
C.E. Creffield,
Phys. Rev. B {\bf 67}, 165301 (2003).

\bibitem{Eckardt2005} 
A. Eckardt, C. Weiss, and M. Holthaus,
Phys. Rev. Lett. {\bf 95}, 260404 (2005).

\bibitem{Lignier2007} 
H. Lignier, C. Sias, D. Ciampini, Y. Singh, 
A. Zenesini, O. Morsch, and E. Arimondo,
Phys. Rev. Lett. {\bf 99}, 220403 (2007).

\bibitem{Kierig2008}
E. Kierig, U. Schnorrberger, A. Schietinger, J. Tomkovic, and M.K. Oberthaler, 
Phys. Rev. Lett. {\bf 100}, 190405 (2008).

\bibitem{Eckardt2009} 
A. Eckardt, M. Holthaus, H. Lignier, A. Zenesini, D. Ciampini, O. Morsch, 
and E. Arimondo, Phys. Rev. A {\bf 79}, 013611 (2009).

\bibitem{Sierra2015} 
C.E. Creffield and G. Sierra,
Phys. Rev. A {\bf 91}, 063608 (2015).

\bibitem{Kitagawa2010} 
T. Kitagawa, E. Berg, M. Rudner, and E. Demler, 
Phys. Rev. B {\bf 82}, 235114 (2010).

\bibitem{Lindner2011} 
N. Lindner, G. Refael, and V. Galitski, 
Nat. Phys. {\bf 490}, (2011).

\bibitem{Wilczek2013} 
F. Wilczek, Phys. Rev. Lett. {\bf 111}, 250402 (2013).

\bibitem{Watanabe2015} 
H. Watanabe and M. Oshikawa, 
Phys. Rev. Lett. {\bf 114}, 251603 (2015).

\bibitem{Choi2017} 
S. Choi {\it et al.}, 
Nature {\bf 543}, 221 (2017).

\bibitem{Zhang2017} 
J. Zhang {\it et al.},
Nature {\bf 543}, 217 (2017).

\bibitem{Russomanno2017} 
A. Russomanno, F. Iemini, M. Dalmonte, and R. Fazio,
Phys. Rev. B {\bf 95}, 214307 (2017).

\bibitem{Yao2017} 
N.Y. Yao, A. C. Potter, I.-D. Potirniche, and A. Vishwanath, 
Phys. Rev. Lett. {\bf 118}, 030401 (2017).

\bibitem{SachaZak2018} 
K. Sacha and J. Zakrzewski,
Rep. Prog. Phys. {\bf 81}, 016401 (2018).

\bibitem{Sacha2018} 
K. Giergiel, A. Kosior, P. Hannaford, and K. Sacha,
Phys. Rev. A {\bf 98}, 013613 (2018).

\bibitem{Goldman2014} 
N. Goldman and J. Dalibard,
Phys. Rev. X {\bf 4}, 031027 (2014).

\bibitem{Holthaus2016} 
M. Holthaus, J. Phys. B {\bf 49}, 013001 (2016).

\bibitem{Russomanno2012} 
A. Russomanno, A. Silva, and G.E. Santoro,
Phys. Rev. Lett. {\bf 109}, 257201 (2012).

\bibitem{Weidinger17}
S.A. Weidinger and M. Knap, Sci. Rep. {\bf 7}, 45382 (2017).

\bibitem{Herrmann18}
A. Herrmann, Y. Murakami, M. Eckstein, and P. Werner, 
Europhys. Lett. {\bf 120}, 57001 (2018).

\bibitem{Yuzbashyan18}
E.A. Yuzbashyan, Ann. Phys. {\bf 392}, 323 (2018).

\bibitem{Sinitsyn18}
N.A. Sinitsyn, E.A. Yuzbashyan, V.Y. Chernyak, A. Patra, and C. Sun, 
Phys. Rev. Lett. {\bf 120}, 190402 (2018).

\bibitem{Korepin1993} 
V.E. Korepin, N. M. Bogoliubov, A. G. Izergin,
{\it Quantum inverse scattering method and correlation functions} 
(Cambridge, Cambridge University Press, 1993).

\bibitem{Mussardo10}
G. Mussardo, {\it Statistical field theory: an introduction 
to exactly solved models in statistical physics} 
(Oxford, Oxford University Press, 2010).

\bibitem{Komnik16}
A. Komnik and M. Thorwart, Eur. Phys. J. B {\bf 89}, 244 (2016).

\bibitem{LiebLiniger1963} 
E.H. Lieb and W. Liniger,
Phys. Rev. {\bf 130}, 1605 (1963).

\bibitem{Yang1969} 
C.N. Yang and C.P. Yang,
J. Math. Phys. {\bf 10}, 1115 (1969).

\bibitem{Yurovsky2008} 
V.A. Yurovsky, M. Olshanii, and D.S. Weiss,
Adv. At. Mol. Opt. Phys. {\bf 55}, 61 (208).

\bibitem{Bouchoule2009} 
I. Bouchoule, N.J. van Druten, and C.I. Westbrook, 
in {\it Atom Chips} (eds J. Reichel, and V. Vuletic) 331-363 (Wiley, 2010). 

\bibitem{Cazalilla2011} 
M.A. Cazalilla, R. Citro, T. Giamarchi, E. Orignac, and M. Rigol,
Rev. Mod. Phys. {\bf 83}, 1405 (2011).

\bibitem{Chen}
H.H. Chen and C.S. Liu, Phys. Rev. Lett. {\bf 37}, 693 (1976).

\bibitem{Ablowitz2004} 
See Appendix D in M.J. Ablowitz, B. Prinari, and A.D. Trubatch, 
{\it Discrete and Continuous Nonlinear Schr\"odinger Systems}
(Cambridge, Cambridge University Press, 2004).

\bibitem{Berry1978} 
M.V. Berry and N.L. Balazs,
Am. J. Phys. {\bf 47}, 264 (1979).

\bibitem{Rau1996} 
A.R.P. Rau and K. Unnikrishnan,
Phys. Lett. A {\bf 222}, 304 (1996).

\bibitem{Guedes2001} 
I. Guedes, Phys. Rev. A {\bf 63}, 034102 (2001).

\bibitem{Feng2001} 
M. Feng,
Phys. Rev. A {\bf 64}, 034101 (2001).

\bibitem{Blanes2009} 
S. Blanes, F. Casas, J.A. Oteo, and J. Ros,
Phys. Rep. {\bf 470}, 151 (2009).

\bibitem{Griffith} 
D.J. Griffiths,
{\it Introduction to Quantum Mechanics}
(Upper Saddle River, NJ: Pearson Prentice Hall, 2005).

\bibitem{Bauch1985} 
D. Bauch,
Nuovo Cimento B {\bf 85}, 118 (1985).

\bibitem{Andreata2004} 
M.A. Andreata and V. V. Dodonov, J. Phys. A {\bf 37}, 2423 (2004); 
J. Rus. Las. Res. {\bf 35}, 39 (2014).

\end{thebibliography}
\end{document}